\documentclass[onefignum,onetabnum]{siamart171218}

\usepackage{lipsum}
\usepackage{amsfonts}
\usepackage{graphicx}
\usepackage{epstopdf}
\usepackage{algorithmic}
\ifpdf
  \DeclareGraphicsExtensions{.eps,.pdf,.png,.jpg}
\else
  \DeclareGraphicsExtensions{.eps}
\fi

\def\xx{{\bf x}}
\def\vv{{\bf v}}

\newsiamremark{remark}{Remark}
\newsiamremark{hypothesis}{Hypothesis}
\crefname{hypothesis}{Hypothesis}{Hypotheses}
\newsiamthm{claim}{Claim}

\headers{A buckling instability}{S. Bachmann , R. Froese, and E. N. Cytrynbaum}

\title{A buckling instability and its influence on microtubule orientation in plant cells}

\author{Sven Bachmann\thanks{Department of Mathematics, The University of British Columbia, Vancouver, BC V6T 1Z2, Canada
  (\email{cytryn@math.ubc.ca})}
\and Richard Froese\footnotemark[1]
\and Eric N. Cytrynbaum\footnotemark[1]}

\usepackage{amsopn}

\makeatletter
\newcommand*{\addFileDependency}[1]{
  \typeout{(#1)}
  \@addtofilelist{#1}
  \IfFileExists{#1}{}{\typeout{No file #1.}}
}
\makeatother

\ifpdf
\hypersetup{
  pdftitle={PlantMTMechanics},
  pdfauthor={S. Bachmann , R. Froese, and E. N. Cytrynbaum}
}
\fi

\begin{document}

\maketitle

\begin{abstract}
In growing plant cells, parallel ordering of microtubules (MTs) along the inner surface of the cell membrane influences the direction of cell expansion and thereby plant morphology. For correct expansion of organs that primarily grow by elongating, such as roots and stems, MTs must bend in the high-curvature direction along the cylindrically shaped cell membrane in order to form the required circumferential arrays. Computational studies, which have recapitulated the self-organization of these arrays, ignored MT mechanics and assumed MTs follow geodesics of the cell surface. Here, we show, through analysis of a derived Euler-Lagrange equation, that an elastic MT constrained to a cylindrical surface will deflect away from geodesics and toward low curvature directions to minimize bending energy. This occurs when the curvature of the cell surface is relatively high for a given anchor density. In the limit of infinite anchor density, MTs always follow geodesics. We compare our analytical predictions to measured curvatures and anchor densities and find that the regime in which cells are forming these cortical arrays straddles the region of parameter space in which arrays must form under the antagonistic influence of this mechanically induced deflection. Although this introduces a potential obstacle to forming circumferentially orientated arrays that needs to be accounted for in the models, it also raises the question of whether plants use this mechanical phenomenon to regulate array orientation. The model also constitutes an elegant generalization of the classical Euler-bucking instability along with an intrinsic unfolding of the associated pitchfork bifurcation.
\end{abstract}


\section{Introduction}

Our understanding of the molecular basis of organismal development and morphology has been accelerating over the last couple decades under the influence of new techniques in molecular biology coupled with increasingly powerful theoretical and computational methods. One area that has benefited recently from these advances is plant development. Plant growth is heavily dependent on the manner in which cells deposit cellulose to form the cell wall because cell expansion occurs most easily in the direction perpendicular to the orientation of cellulose fibres \cite{Hamant2010}. Cellulose deposition is influenced, in turn, by ordered arrays of microtubules (MTs) that adhere to the inner surface of the cell membrane by anchoring to the cell wall \cite{Elliott2018}. Remarkably, unlike animal MT organization that is largely driven by centrosomes, for example, as seen in the mitotic spindle, these plant cortical arrays self-organize by a combination of cell geometry, distributed nucleation, and MT-MT interactions. 

When a growing MT encounters an existing MT, the interaction can lead to one of three outcomes that together have been proposed to explain ordering: induced catastrophe, entrainment/zippering, or cross-over and subsequent severing \cite{Dixit2004}. Early computational and mathematical analysis demonstrated that models of cortical MTs on a two dimensional surface could self-organize into highly ordered arrays driven only by these MT-MT interactions \cite{Baulin2007, Allard2010MBC, Tindemans2010, Eren2010, Hawkins2010, Shi2010, Deinum2011}. Simulating such arrays on cylinders with inaccessible end-caps, the arrays align circumferentially as seen in growing plant cells \cite{Allard2010MBC, Eren2010}.  At the level of the whole cell, extensions of these models have been implemented in more complex geometries such as polyhedral domains \cite{Ambrose2011} and more recently in finely resolved triangulations of actual cells \cite{Chakrabortty2018}. Another recent model examined the dynamic relocalization of MTs from cytosol to cortex in abstract 3D geometries to assay the influence of cell shape on MT distribution and orientation, in the absence of MT-MT interactions \cite{Mirabet2018}. These whole-cell models have helped clarify how array organization is globally coordinated across geometrically complicated domains and in some cases how MT-associated proteins can regulate that coordination \cite{Ambrose2011, Chakrabortty2018}.

In all of these models, MT growth was assumed to follow geodesics of the cell surface and the influence of cell wall curvature deflecting MTs through bending in the tangent plane was ignored (in one stochastic model, MTs ought to follow geodesics on average \cite{Mirabet2018}). Although some authors have given brief justification for this simplifying assumption \cite{Allard2010MBC, Tindemans2010, Tindemans2014}, the issue has never been properly addressed in the context of this type of model. 

The cortical-array organization of MTs had been studied in earlier work using {\it in vitro} experiments in fabricated cylindrical geometries containing free-growing MTs and, to understand these experiments, using thin-rod elasticity theory applied to theoretical rods in an analogous geometry \cite{Lagomarsino2007}. Although the theory was accurate in its prediction of the MT configurations seen in the experiments, neither the experiments nor the theory were consistent with the configurations observed in analogous cylindrically shaped growing plant cells. As later pointed out, cortical MTs in plants are not free to move laterally on the cell membrane because they are anchored along their lengths \cite{Tindemans2014}. In the latter article, the authors justified their own and others' assumption that elasticity of MTs did not play a role in determining MT growth direction by invoking the stiffness of MTs and the frequency of anchoring. Essentially, they claimed that for unanchored MT tip lengths that are small compared to the local radii of curvature of the cell wall, MT deflection away from geodesics, induced by minimizing bending energy, would be negligible. While this claim is true, a rigorous treatment of this claim has never been given. Nor is it clear that plant cells are in the alleged scaling regime. The relevant radii of curvature of cell walls have been measured to be in the range of $7 \ \mu m \pm 2.5 \mu m$ (mean $\pm$ standard deviation) around longitudinal edges that MT arrays are seen to circumvent \cite{Ambrose2011}. Distributions of unanchored MT tip lengths have also been measured and average around $3 \ \mu m$ with tails extending up to $8 \ \mu m$ \cite{Ambrose2008, Allard2010BiophysJ}. Although, on average, the reported unanchored lengths are smaller than the reported curvatures, the distributions are overlapping which makes predictions about the untested influence of mechanics on array orientation difficult to assess.

In this paper, we address this question directly: To what extent does minimization of bending energy allow the curvature of the cell wall to deflect the unanchored tips of cortical MTs away from geodesic paths?

\section{Model equation derivation}

We consider a single microtubule (MT) anchored at the inner surface of a cell membrane. In a plant cell, the cell wall is much more rigid than a MT so we assume the cell-wall shape is a fixed cylinder and the MT shape is determined by a process of bending-energy minimization. Anchoring proteins along the length of the MT keep it tight to the membrane but, as growth occurs at the plus end, a short length of the MT at that end, which we call the tip, is free to fluctuate thermally.  While free, this tip can explore the surface of the membrane, and to some extent the cytosol, before getting pinned down by an anchor somewhere along the free length. We examine the problem of minimizing the bending energy of a single MT tip and consider implications for the organization of cortical MT arrays. 

\begin{figure}
      \centering
      \includegraphics[width=7cm]{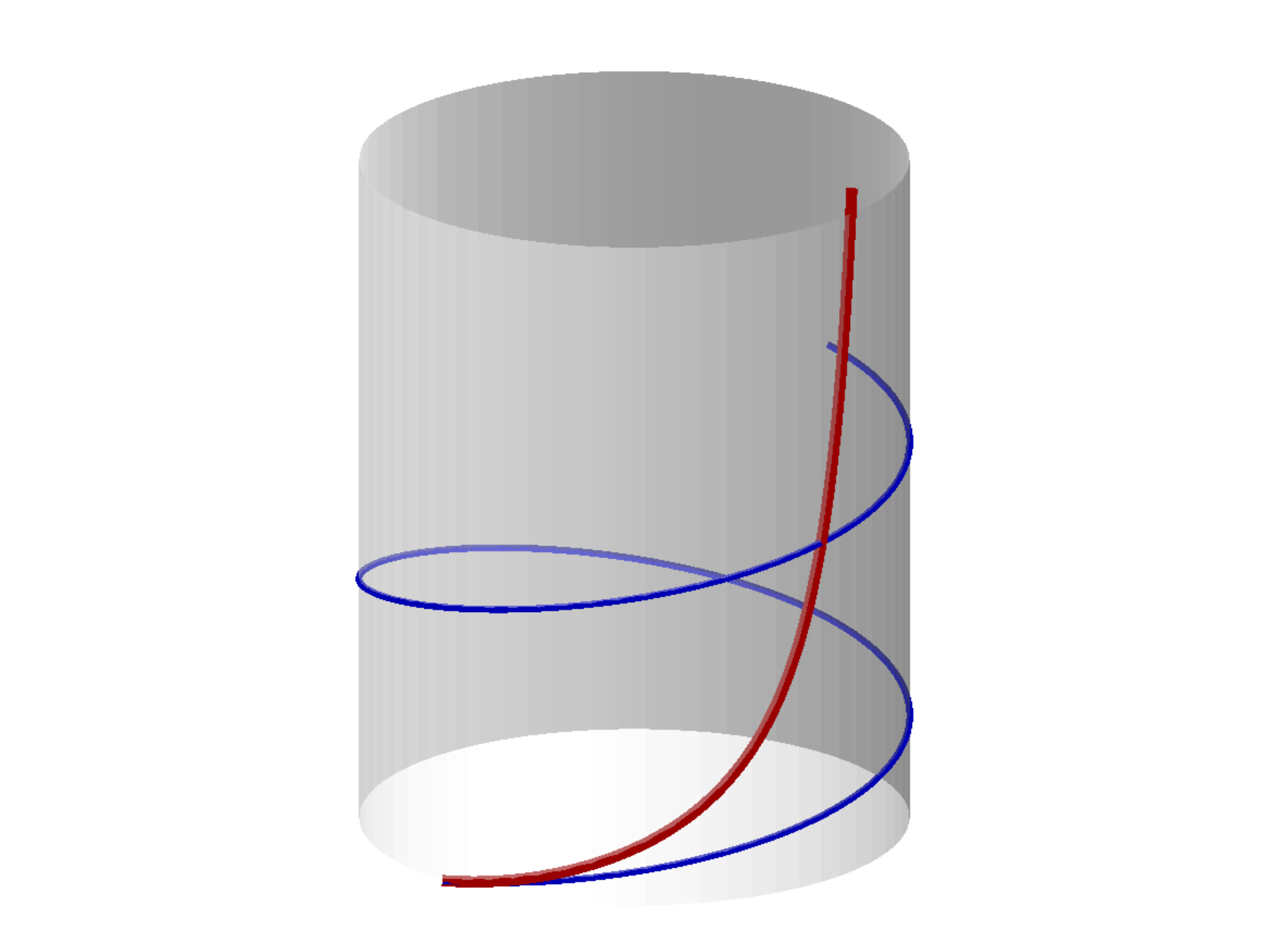} 
      \caption{An illustration of the expected behaviour of a MT tip on a curved surface. The thinner (blue) curve is a geodesic (helix) and the thicker (red) curve is a MT tip decreasing its total bending energy by deflecting away from the geodesic and heading toward the lower curvature vertical orientation. The plus ends, where the MTs grow, are at the top. The foremost anchor on the MT, which defines the other end of the tip, is at the bottom. The rest of the MT is not shown. The blue and red curves share the same position and angle at the anchor point.}
      \label{fig:Cartoon}
   \end{figure}

On a flat surface, the path of MT growth should be straight. When a surface has curvature, the MT bends. Previous work on MTs growing in such conditions assumed that MTs follow geodesics of the surface, as described above. However, when the surface has differing principal curvatures, we expect that the equilibrium configuration is one that deviates from a geodesic so as to turn the MT tip closer to the principal curvature direction with lower curvature (see Fig.~\ref{fig:Cartoon} for illustration). More precisely, the shape of the MT between its plus end and the nearest anchor point, at the base of the tip, should be such that it minimizes its total bending energy given by the scalar curvature
\begin{equation}
{\cal K}[\varphi] = \int_0^L ((\cos\varphi)^4+\dot \varphi^2)\; ds,
\label{eq:energyfunctional}
\end{equation}
under appropriate boundary conditions. Here, the shape of the MT tip is described by the function $\varphi(s)$ which gives the angle formed between the circumferential direction of the cylinder and the tangent vector to the MT at a distance $s$ from the anchored end of the MT tip.

In Appendix \ref{sec:ELderivation}, we derive the functional~\eqref{eq:energyfunctional} and the corresponding Euler-Lagrange equation:
\begin{equation}
\frac{1}{2}\ddot \varphi + (\cos\varphi)^3\sin\varphi =0, \quad \varphi(0)=\varphi_0, \quad \dot\varphi(L)=0.
\label{eq:EL}
\end{equation}
An abridged derivation of this equation has been published previously with different boundary conditions to describe free MTs in constrained geometries \cite{Lagomarsino2007}.

\subsubsection*{Hamilton's equations}

Analysing the problem in the phase plane is informative so we recast the Euler-Lagrange equation~\eqref{eq:EL} as a Hamiltonian system. Using standard methods, we derive the Hamiltonian
\[
H(\varphi,p) = \frac{1}{4}p^2 - (\cos\varphi)^4.
\]
and the 
first order Hamiltonian system
\begin{align}
\begin{split}
\dot\varphi &= \frac{p}{2},\\
\dot p &= - 4(\cos \varphi)^3\sin(\varphi).
\label{eq:cylsys}
\end{split}
\end{align}

\begin{figure}
      \centering
      \includegraphics[width=8cm]{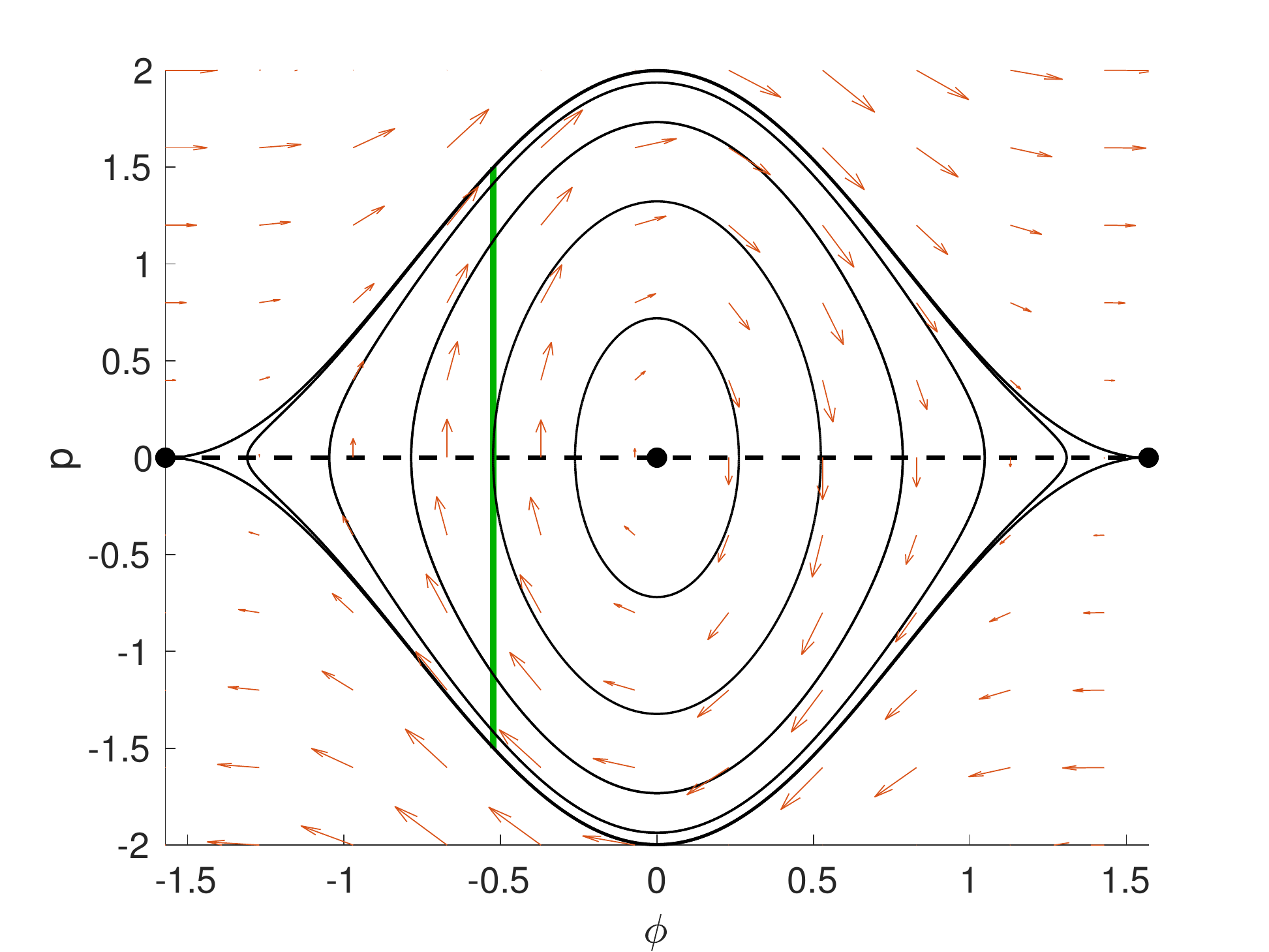} 
      \caption{Constructing solutions $(\varphi(s),p(s))$ to the boundary value problem (BVP) for the cylinder case in the $\varphi-p$ phase plane. Solutions must start at \(\varphi(0)=\varphi_0\) (vertical green line segment) and end on \(p(L)=0\) (dashed black line segment). As \(s\) increases, each left boundary condition (at $s=0$) on the green line flows according to the vector field defined by \eqref{eq:cylsys} (red arrows). A few level curves of the Hamiltonian are shown (black curves).}
      \label{fig:PhasePlane}
   \end{figure}

We illustrate the Hamiltonian vector field and some of its integral curves in Fig.~\ref{fig:PhasePlane}. Solutions of the boundary value problem (BVP) $\varphi$ 
correspond to integral curves starting along the vertical line $\varphi=\varphi_0$ and ending on the horizontal line \(p=0\) when \(s=L\). This being a Hamiltonian system, solutions stay on a level curve of \(H\).

\section{Numerical methods}

All numerical calculations were carried out using \textsc{MATLAB}\textsuperscript{\textregistered}. To solve the EL equations \eqref{eq:cylsys} with boundary conditions $\varphi(0)=\varphi_0$ and $p(L)=0$ as required for Fig.~\ref{fig:PhasePlanePhi00}, we integrated a family of initial conditions with $\varphi(0)=\varphi_0$ and $-2(\cos\varphi_0)^2\le p(0) \le 2(\cos\varphi_0)^2$ and selected the solution having $p(L)$ closest to zero.
For Fig.~\ref{fig:Segments}, we integrated forward in $s$ starting with two left boundary conditions $(\varphi(0),p(0))=(\varphi_0,0)$ and $(\varphi(0),p(0))=(\varphi_0,2)$ using \textsc{Matlab}\textsuperscript{\textregistered}'s fourth order Runge-Kutta scheme \textsc{ode45} and used a bisection method on $p(0)$ to approximate the solution satisfying $p(L;p(0))=0$. To generate the bifurcation diagrams, we integrated equation \eqref{eq:cylsys} over a two-parameter family of $p(0)$ and $L$ values and approximated the diagram with a contour plot of $p(L)=0$. 

\section{Results}

\subsubsection*{MTs growing parallel to the cylinder axis}

The steady state solution $(\varphi(s), p(s)) = (\pi/2,0)$ corresponds to a MT growing parallel to the axis of the cylinder and satisfies the right boundary condition $p(L) = 0$ for all $L$. Since the curvature functional~\eqref{eq:energyfunctional} is non-negative in general and vanishes in this steady state, this solution is a global minimizer of ${\cal K}$ and therefore a stable solution. This is, of course, clear from the physical point of view since any bending away from the axial direction increases both the cylindrical bending (the $\cos$ term in $\mathcal{K}[\varphi]$) and the bending of the MT in the tangent plane (the $\dot\varphi$ term). The symmetric solution for $\varphi_0=-\pi/2$ has similar properties but points in the opposite direction.

As suggested by the phase portrait, there is also a non-constant solution of the Euler-Lagrange equation associated with $\varphi_0 = \pi/2$: it is the heteroclinic connection between \((\pi/2,0)\) and \((-\pi/2,0)\). However, for any finite $L$ it is not a solution of the minimization problem since it reaches $p=0$ only for $L\to\infty$: This heteroclinic corresponds to an infinitely long MT that comes from infinitely far down the cylinder, undergoes a half turn and heads back down the cylinder. The reverse heteroclinic is the mirror image solution on the cylinder.

\subsubsection*{A buckling instability for circumferentially growing MTs}
Solutions having \(\varphi_0=0\) have a richer structure. They correspond to a MT initially oriented circumferential to the cylinder, namely in the direction of maximal curvature on the cell. The constant \(\varphi(s)=0, p(s)=0\) is a solution for any value of \(L\). On the cylinder, this solution is an arc of a circle around the circumference of the cylinder. 

Although intuitively a circumferential solution might seem to maximize the curvature, the situation is more subtle. In order for a MT tip, clamped in the circumferential direction at one end, to reduce its total bending energy by exploiting the flat axial direction along the cylinder, it must bend away from its initial direction. This induces an additional bending in the tangent plane on top of to the bending normal to the cell surface which is necessary to stay on that surface. Hence, it is natural to expect a transition in stability as a function of the length $L$, with the MT preferring the circumferential direction for small $L$, but bending away from it and towards the axial direction whenever $L$ is large enough to ensure a sufficiently long interval of MT heading in the flat direction.

\begin{figure}[h]
      \centering
      \includegraphics[width=8cm]{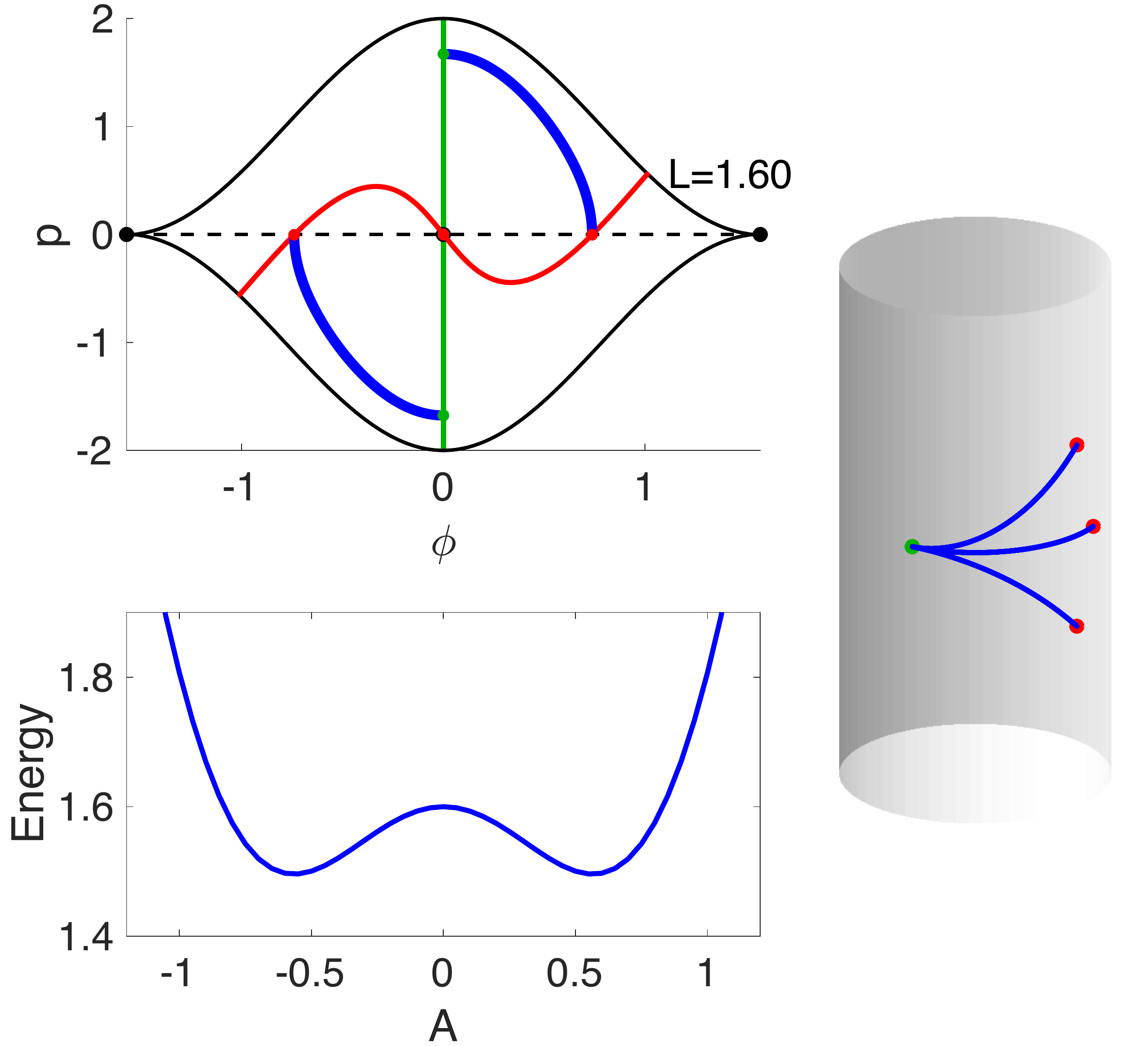} 
      \caption{Numerically calculated solutions to the BVP in the phase plane (top left), on the cylinder (right), and their bending energy (bottom left). Top left: The solid vertical (green) line and the dashed horizontal line represent boundary conditions as in Fig.~\ref{fig:PhasePlane}. The sinusoidal-like (red) curve shows the result of flowing the entire family of ICs (vertical green line segment) forward from \(s=0\) to \(s=L\) for \(L=1.6\). The solutions to the BVP are drawn as thick (blue) curves connecting the dots at \((\varphi(0),p(0))\) and \((\varphi(L),p(L))\). The case of \(L=1.6\) is just above the first pitchfork bifurcation at \(L_0\). Bottom left: The energy of a MT with shape given by \(A\sin(\sqrt2s)\). Note the maximum at \(A=0\) indicating that the circumferential solution is unstable and the minima close to 0.6 corresponding to the stable solutions that bend up/down the cylinder wall.}
      \label{fig:PhasePlanePhi00}
   \end{figure}

In Appendix \ref{sec:bucklingDetails}, we show that for $L<L_0$ where $L_0=\pi/2\sqrt{2}\approx 1.1$, there is only the circumferential solution \(\varphi(s)=0, \ p(s)=0\) to the EL equation with $\varphi_0=0$. 
Furthermore, we find that at $L=L_0$, there is a supercritical pitchfork bifurcation through which a symmetric pair of stable solutions appear and the zero solution goes unstable. Numerical approximations of the three solutions to the EL equation for a value of $L$ just above the bifurcation are shown in the top left and right panels of Fig.~\ref{fig:PhasePlanePhi00}. The former shows the solutions in the phase plane and the latter shows them as rods on the cylinder.

We also derive an implicit expression for the angular direction of these solutions at the end of the tip, $\varphi_f(L)$:
\begin{equation*}
L = \int_{0}^{\varphi_f} \dfrac{1}{\sqrt{(\cos\varphi)^4-(\cos\varphi_f)^4}}d\varphi
\end{equation*}
which can be inverted numerically to get the shape of the emerging branches. As $L$ increases, $\varphi(L)$ approaches $\pi/2$ meaning that this solution bends closer to the axial direction. 

In addition to the pitchfork bifurcation at $L_0$, for each $n=1,2,3...$, there is a pitchfork bifurcation at 
\begin{equation}
L_n = \frac{(1+2n)\pi}{2\sqrt{2}}.
\end{equation}

The stability of the zero solution can be understood by analyzing the energy functional in a neighbourhood around it. We find that the zero solution is a minimum of the energy functional  for $L<L_0$ and hence is stable there. At $L=L_0$, a single mode goes unstable and for all $L>L_0$ the zero solution remains unstable. This analysis is explained in Appendix \ref{sec:bucklingDetails}.  To confirm that the branching solution that appears at $L_0$ is stable, we evaluate the energy functional numerically along the line defined by $\varphi_u(s)=A\sin(\sqrt{2}s)$ for several values of $L$ just above $L_0$. The mode $\varphi_u$ is the one that goes unstable at $L_0$ and the line spanned by it should be a close approximation of the heteroclinic connecting the zero solution to the branching solution for $L$ above but close to $L_0$. We find a pair of minima for some value of $\pm A$ that gradually move away from the zero solution with increasing $L$. This energy profile is shown in the bottom left panel of Fig.~\ref{fig:PhasePlanePhi00}. Note the maximum at $A=0$ (unstable) and the two minima near $A=\pm0.6$ (stable).

This result corresponds precisely to the intuitive explanation of the bifurcation given above. We now see that the tradeoff of tangential bending close to the anchor point in exchange for less bending farther away reaches a balance at $L_0$, with it being worthwhile above $L_0$ but not below. 

\begin{figure}
      \centering
      \includegraphics[width=8cm]{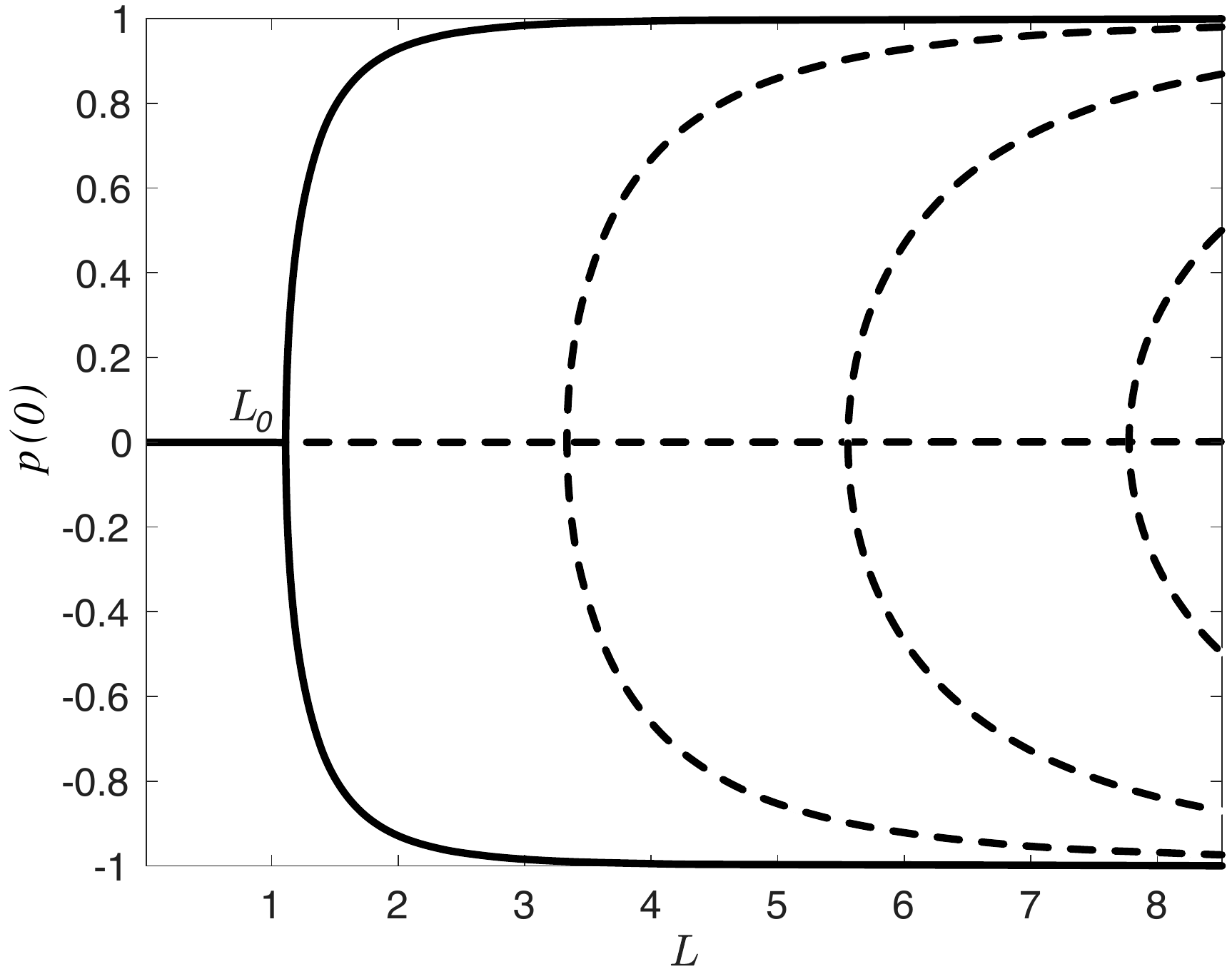} 
\caption{The bifurcation diagram, calculated numerically, showing solutions to the EL equations with \(\varphi_0=0\). Solid curves denote stable solutions and dotted curves denote unstable solutions, determined analytically for the zero solution. The stability of the solutions bifurcating at $L_0$ is inferred from the numerical calculation of the energy for several instances of $L$ (see the bottom left panel of Fig.~\ref{fig:PhasePlanePhi00} for an example).}
\label{fig:pitchfork}
\end{figure}

The bifurcation diagram with stability denoted is outlined in Fig.~\ref{fig:pitchfork}. 

\subsubsection*{Deflection of MTs growing close to circumferentially}

\begin{figure}
      \centering
      \includegraphics[width=8cm]{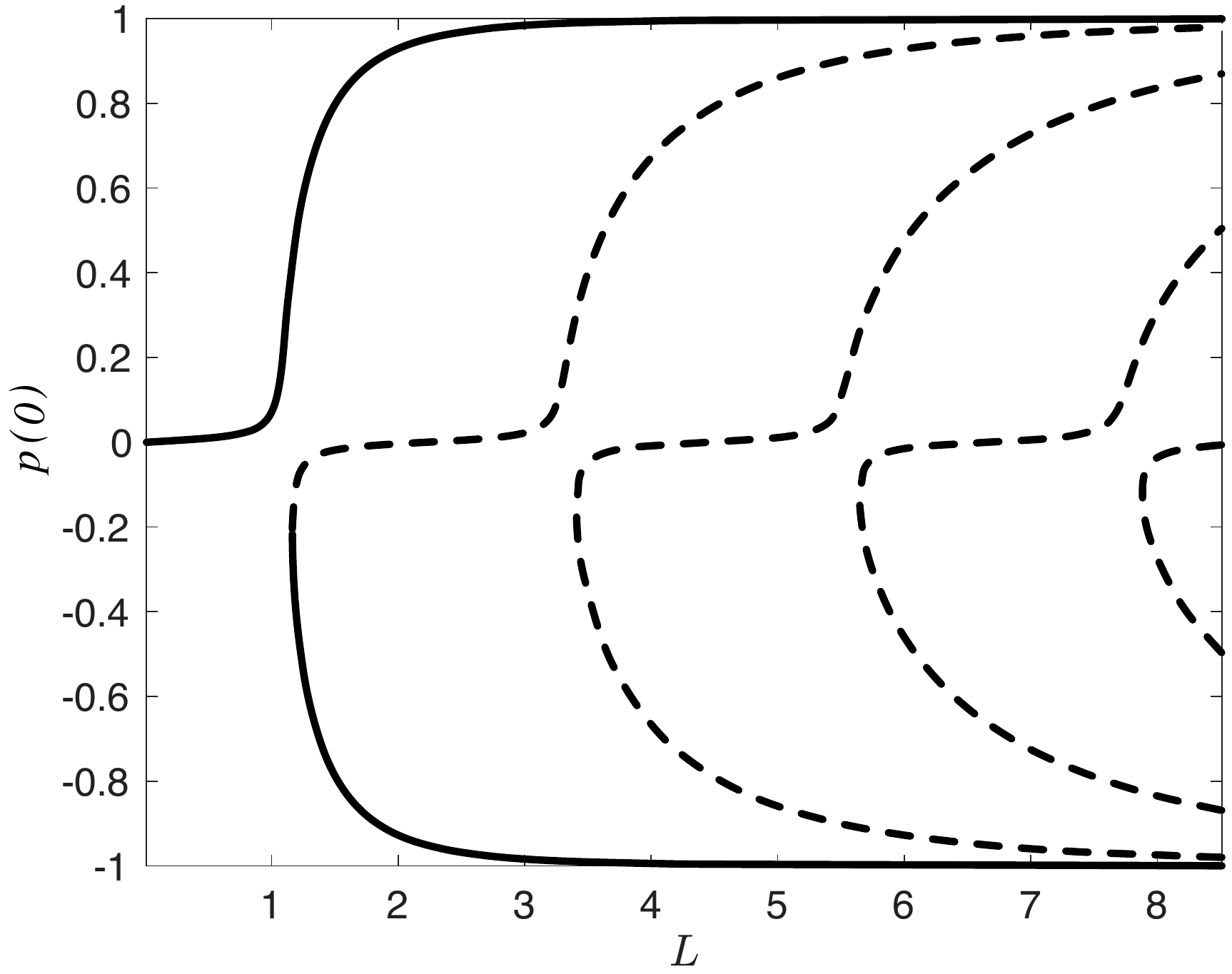} 
\caption{The bifurcation diagram showing solutions to the EL equations with \(\varphi_0=0.008\). We did not independently determine stability either analytically or numerically in this case but instead infer it from the stability for the symmetric case ($\varphi_0=0$) with solid curves denoting stable solutions and dotted curves denoting unstable solutions. }
\label{fig:saddlenode}
\end{figure}

A complete bifurcation analysis as carried out for the \(\varphi_0=0\) case is not as simple for the case when \(\varphi_0\ne0\). However, we can still approximately substantiate the main message from the previous section: if MTs anchor frequently to the cell wall and the lengths of all segments are short then they grow nearly along geodesics. With \(\varphi_0\ne0\), symmetry is broken and the pitchfork unfolds into a stable steady state and a saddle-node pair. Fig.~\ref{fig:saddlenode} shows the numerically calculated bifurcation diagram for the case of $\varphi_0=0.008$. 
The biologically relevant feature is that for $L$ below $L_0$, $p(0)$, and hence $\varphi(L)$, is still close to zero for much of that interval. However, even for this small value of $\varphi_0$, below but close to $L_0$ there is already a significant deflection. 

It has been commonly assumed in previous modelling work that higher frequency anchoring reduces the total deflection away from circumferential. The bifurcation diagram in Fig.~\ref{fig:saddlenode} shows that the deflection approaches zero as $L$ approaches zero, a necessary but insufficient condition to justify this assumption. We must further ask whether these almost-zero deflections may add up to something significant over many anchored segments.

We compare the total deflection when the MT attaches only at $L$ with the total deflection when it attaches sequentially $n$ times to the surface with equal spacing $\ell = L/n$. We enforce a no-kink condition at each attachment point. For $1\leq j\leq n$, let $\varphi_j$ be the solution of the Euler-Lagrange equation on $[0,\ell]$ with boundary conditions $\dot\varphi_j(\ell) = 0$ and $\varphi_j(0) = \varphi_{j-1}(\ell)$. Here we set $\varphi_1(0) = \varphi_0$. The total angular change is then given by the telescopic sum
\begin{equation*}
\Delta_L(\varphi_0,n) = \sum_{j=1}^{n}(\varphi_j(\ell) - \varphi_j(0)).
\end{equation*}
In this notation, $\Delta_L(\varphi_0,1)$ corresponds to the scenario discussed above of a tip growing to length $L$ and being anchored only at the end without intermediate anchoring while $\Delta_L(\varphi_0,n)$ corresponds to the scenario in which there are $n$ equally spaced anchors.
In Appendix \ref{sec:DeltaBoundProof}, we prove that
\begin{equation}
\vert \Delta_L(\varphi_0,n) \vert\leq \frac{L^2}{n}.
\label{eq:DeltaBound}
\end{equation}

\begin{figure}
      \centering
      \includegraphics[width=8cm]{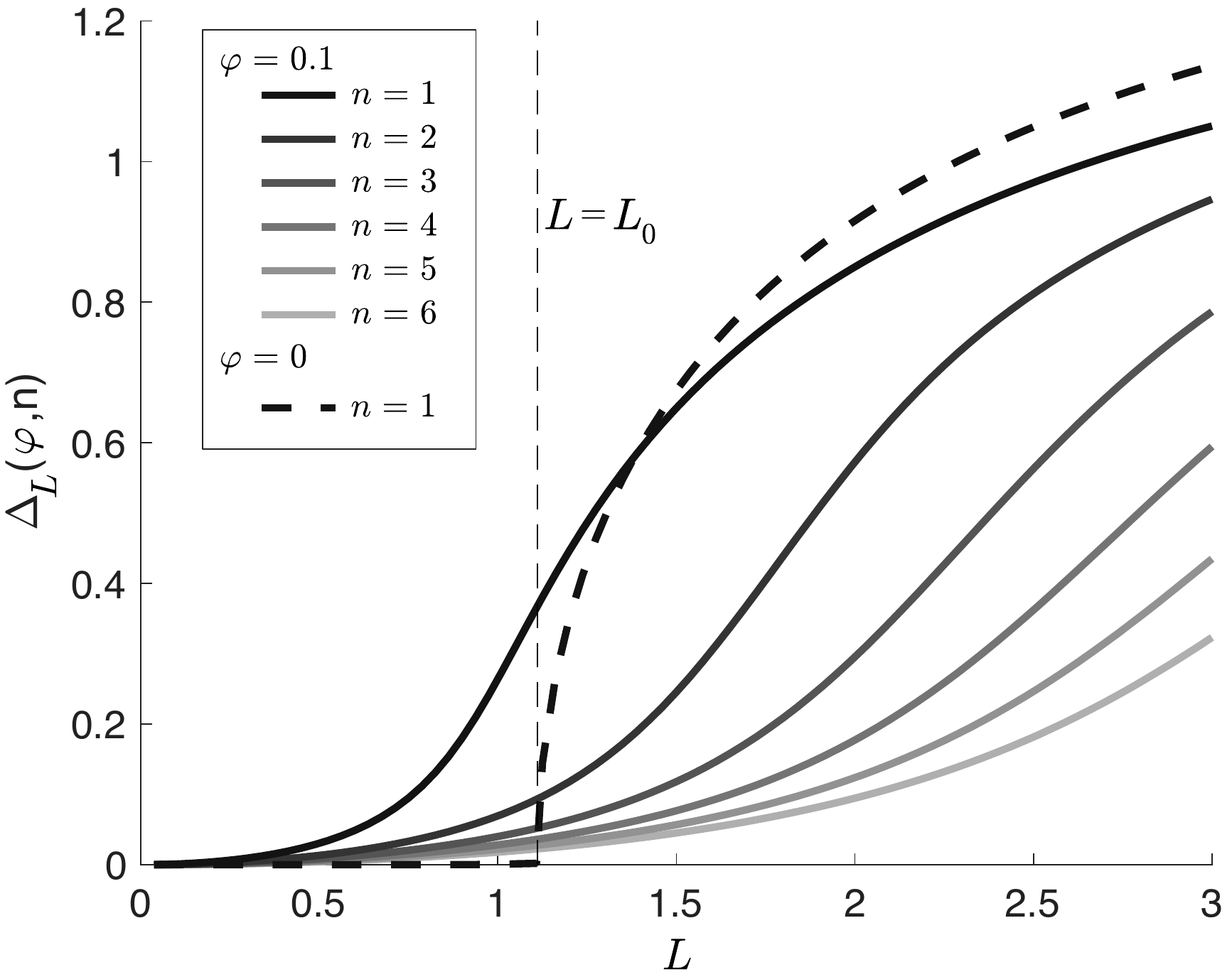}
      \caption{Deflection of nearly circumferentially growing MTs. A perfectly circumferentially growing MT does not deflect when the tip length is below $L_0$ but above that critical length it deflects dramatically (dashed curve). A nearly circumferentially growing MT with an initial angle of $\varphi_0=0.1$ ($\sim6^\circ$) will deflect significantly well below $L_0$ (darkest solid curve). For the same initial angle and the same total length of MT, anchoring the MT incrementally so that the tip length at each incremental anchor is shorter ($L/n$) leads to a marked decrease in total deflection (lighter solid curves).}
      \label{fig:Segments}
   \end{figure}

Let us explain how Eq.~\eqref{eq:DeltaBound} supports the claim that high anchor frequency leads to near-geodesic MT growth for $\varphi_0 \ne 0$. From \eqref{eq:DeltaBound} it is clear that for any $L$
\begin{equation*}
\vert \Delta_L(\varphi_0,n) \vert \leq \Delta_L(\varphi_0,1),
\end{equation*}
for $n$ large enough. In fact, we calculated $\Delta_L(\varphi_0,n)$ numerically for $\varphi_0=0.1$ and several values of $n$ (see Fig.~\ref{fig:Segments})
and found that $\Delta_L(\varphi_0,n)$ appears to be a decreasing sequence in $n$ for any fixed $L$. Thus, more anchors means less deflection. Furthermore, Eq.~\eqref{eq:DeltaBound} implies
\begin{equation*}
\lim_{n\to\infty}\Delta_L(\varphi_0,n) = 0.
\end{equation*}
From this, we reach the same conclusion as in the case $\varphi_0=0$, although not as sharply. If $n$ is large enough, then the total angular change approaches zero for any fixed $L$. This conclusion brings us halfway to justifying the assumption that cortical MTs follow geodesics. The one issue to resolve, which we address in next section, is how long are the MT tip lengths in real cells. 

\subsection*{Curvature and anchoring data suggest significant MT deflection}

Recall that the length $L$ in our model is the nondimensional tip length with the background scale being the radius of the cylinder. To extract this from data, we need measurements of tip lengths and radii of curvature from actual cells. Tip lengths have been measured previously \cite{Ambrose2008} and analyzed using a growth and anchoring model \cite{Allard2010BiophysJ}. Typical radii of curvature have also been measured \cite{Ambrose2011}. We used the mean ($7 \mu$m), SEM ($0.2 \mu$m), and sample size ($n=154$) of radii of curvature for longitudinal edges from root epidermal division zone cells from Fig.~3c of \cite{Ambrose2011} to reconstruct an approximation of the original distribution in the data. Then, combining that with the actual distribution of tip lengths in wildtype cotyledon epidermal cells shown in Fig.~4 of \cite{Ambrose2008} ($4.2 \pm 1.6 \mu$m), we generated a synthetic data set of nondimensional tip lengths by randomly sampling both distributions and calculating their ratio. Obviously, no MT whose tip length was measured in \cite{Ambrose2008} ever crossed the longitudinal edges from which the curvatures were measured in \cite{Ambrose2011} but these distributions represent reasonable scales for the required quantities. 

\begin{figure}
      \centering
      \includegraphics[width=8cm]{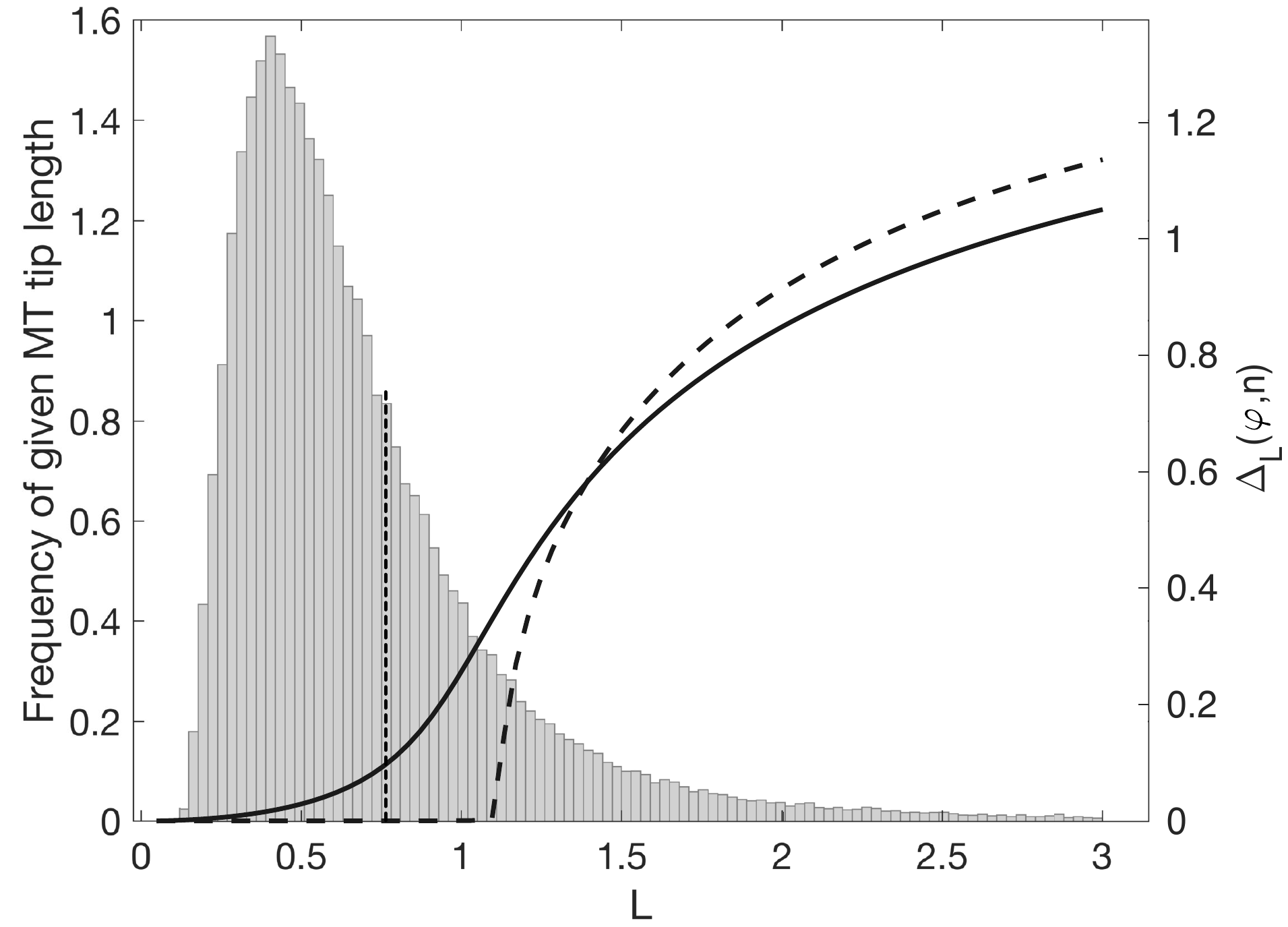}
      \caption{A histogram of nondimensional tip lengths generated from measured distributions of tip lengths and radii of curvature. Superimposed on the histogram are the deflection graphs for $\varphi_0=0$ (dashed, with a kink at the critical value) and $\varphi_0=0.1$ (solid) from Fig.~\ref{fig:Segments}. If we consider a deflection of 0.1 ($\sim6^\circ$) or larger for a MT starting at an angle of 0.1 to be a substantive deflection, roughly one third of the nondimensional tip lengths (to the left of the vertical dotted line) fall into this category.}
      \label{fig:data}
   \end{figure}

The resulting distribution is shown as a histogram in Fig.~\ref{fig:data}.
As explained in the caption, although many MTs are predicted to stay close to geodesics, roughly one third of the ``synthetic'' MT tips are predicted by the model to undergo substantial deflection. This raises serious concerns for the assumption commonly made that cortical MTs follow geodesics, especially given that the deflection is away from the circumferential direction.

\section{Discussion}

The issue of how MT mechanics influences the organization of cortical arrays in plant cells has remained largely unexplored until now. Here, we found that MTs with sufficiently long unanchored lengths at their plus ends deflect from high to low curvature directions on the cortical surface of the cell. Given the important role of circumferential orientation of cortical MT arrays for cellulose deposition and hence plant growth and morphology, the density of anchors is thus a critical parameter that plant cells must control. 

Specifically, we found that the free tip of a MT with its anchored end fixed exactly in the circumferential direction deflects significantly towards the axial direction when the tip length is larger than a critical length of $L_0 \times R$ where $L_0=\pi/2\sqrt{2} \approx1.1$ and $R$ is the radius of curvature in the circumferential direction. This deflection also occurs for MTs whose anchored ends are fixed close to the circumferential direction, even below the critical length. On the other hand, in the limit of short tip lengths, we proved that deflection goes to zero fast enough that, when added up over many anchored segments, the total deflection over a longer length of MT is still negligible.  

Although we have shown that surface curvature can influence individual MT orientation, we have not addressed how this tendency plays out at the level of array organization and orientation tuning. Our analysis of previously published data on tip lengths and curvatures of cell walls indicates that a large fraction of MTs in the simulations of the ordering and orienting of cortical arrays should actually be subject to a deflecting force that is disruptive to circumferential organization. The effect of this force on array orientation needs to be tested in the full context of cortical MT-MT interactions and MT dynamic instability on complex cell geometries. Complicating the story of MT array orientation tuning, other factors have also been found to influence it. In particular, stress in the cell wall has been found to influence the orientation of MTs \cite{hamant2008developmental}, possibly through the tension-dependent modulation of MT dynamic instability parameters which has been observed experimentally \cite{franck2007tension,trushko2013growth} and conjectured to be the link between stress and MT orientation selection \cite{Hamant2019}. All these factors need to be tested in a comprehensive model of array formation and orientation to determine when, where and how each of these factors matter.

Interestingly, our analysis is independent of the flexural rigidity or Young's modulus of the MT. In fact, a dimensional version of the energy functional given by Eq.~\eqref{eq:energyfunctional} would have the flexural rigidity as a coefficient but it would fall out of the Euler-Lagrange equation due to homogeneity of the equation. The flexural rigidity would matter in two circumstances: (1) if we did not assume an infinite stiffness for the cell wall, the ratio of the elastic moduli of the cell wall and the MT would determine their relative deformations; (2) the probability that MT polymerization slows or stalls on a curved surface is dependent on curvature because a MT must deform to allow for the addition of subunits in the manner of an elastic ratchet \cite{mogilner1996cell}. Clearly, the ratio of elastic moduli is finite but large enough that cell wall deformation can be ignored. On the other hand, slowing or stalling of MT growth due to curvature is likely to occur. In the present work, we are focused on those MTs that continue to grow around corners and their speed is not relevant in our model, however, in a full-context model alluded to in the previous paragraph, this effect could be important.

Also relating to the flexural rigidity, one might ask how the energy scale of bending compares to the thermal energy scale. For a typical MT tip length ($4.2 \ \mu$m) and radius of curvature ($7 \ \mu$m), the bending energy is roughly $1000 k_BT$ so we do not expect thermal fluctuations to make a large impact on the results described here.

Although polymerization speed is not a parameter in our model, it does provide for an interesting anchor-density regulatory  mechanism. At slower polymerization speeds, anchor density is predicted to be higher even if the background rate of anchor formation is the same \cite{Allard2010BiophysJ}. 

We have made several simplifying assumptions that may be worth revisiting for the problem we considered here and for those that we cannot consider given these simplifications. For example, in our model, new anchors attach at the very end of a MT tip. In reality, anchors could attach anywhere along the tip length, the impact of which on our quantitative predictions is not entirely clear. Cell geometry is another detail that requires further exploration. Although a cylinder is a reasonable first approximation of the shape of an elongating root cell, these cells tend to have polygonal rather than circular cross sections. Other plant cells, for example leaf pavement cells, have even more complicated geometries and microtubule orientation still appears to play a role in that complexity\cite{Akita2015}. 

Our mathematical analysis generalizes the classical analysis of the buckling of a thin elastic rod under compression. The connection between these two is made more clear by considering an alternate physical scenario that leads to the same mathematical formalism discussed here. Suppose a thin elastic rod is clamped at one end to a flexible sheet and the sheet is slowly curled into a cylinder. If the clamped end is oriented perpendicular to the cylinder's axis, the rod buckles through a pitchfork bifurcation when the product of the curvature of the sheet with the length of the rod is $L_0$, analogous to the classic rod buckling. In addition, this system provides an elegant unfolding of that pitchfork bifurcation when the orientation of the rod is not perpendicular to the cylinder axis.

In light of our findings, a few questions become a priority. Do simulations of cortical arrays still orient correctly under the deflecting influence of cell wall curvature? What is the actual distribution of MT tip lengths as MTs circumvent highly curved regions of the cell surface, especially near sharp edges where CLASP may be modulating deflection by promoting anchoring or where slowing of polymerization might be increasing anchor density? When MT arrays rearrange from a circumferential orientation during growth to more disordered or longitudinally oriented during non-growth phases, could the transition be driven by a regulated decrease in anchor density or increase in polymerization speed? How does cell geometry influence MT array organization in cells with more complicated geometry, for example leaf pavement cells, and is MT mechanics enough to explain the relationship?

\appendix

\section{Derivation of the Euler-Lagrange equation} 
\label{sec:ELderivation}

\subsection*{MTs on a cylindrical cell surface}

We describe a plant cortical MT as a curve embedded within the cell surface, in the case described here, a cylinder. We make a few simplifying physical assumptions that allow us to treat each segment of MT between neighbouring anchor points independently. (1) We assume that anchors always attach the MT to the cell surface at the location of the plus end of a growing MT at the moment it anchors. (2) Upon attachment of an anchor at the plus end, the position and orientation of that point along the MT are henceforth fixed. This is equivalent to saying that anchors act as infinitely stiff positional and torsional springs. (3) As the MT grows beyond a new anchor,  the shape of the new tip is determined by minimizing the total bending energy along the entire length of the MT tip under the assumption that it is on the cell surface at all points along the tip. In the calculations that follow, in light of assumption (2), we are always considering the shape of the MT tip and not the rest of the MT given that behind the foremost anchor the MT shape is fixed. 

We consider curves constrained to lie on the (infinite) cylinder
\[
C = \left\{\left[\begin{matrix}\cos\theta\\  y \\ \sin\theta\\ \end{matrix}\right] : \theta \in [0,2\pi],y\in\mathbb{R} \right\},
\]
where units are chosen so that the cylinder has a (nondimensional) unit radius. All lengths in the model can thus be interpreted dimensionally by multiplying the nondimensional length by the actual radius of the cell in question. 
For a parametrized curve in \(C\) given by  
\[\xx(s) =   \left[\begin{matrix}\cos(\theta(s))\\  y(s) \\ \sin(\theta(s))\\ \end{matrix}\right],\]
the tangent vector $\vv$ is given by
\[
\vv(s) = \dot\xx(s) =   \left[\begin{matrix}-\sin(\theta(s))\\  0 \\ \cos(\theta(s))\\\end{matrix}\right]\dot \theta(s)  + \left[\begin{matrix}0 \\  1 \\  0\\ \end{matrix}\right]\dot y(s)
\]
so that \(\|\vv(s)\|^2 =\dot\theta(s)^2 + \dot y(s)^2\). Thus \(\xx\)
is parametrized by arclength if \(\dot\theta(s)^2 + \dot y(s)^2 = 1\). Without loss of generality, we set \(\theta(0)=y(0)=0\). The tangent vector and hence the curve is uniquely determined by a single function $\varphi(s)$ with the relations
\begin{equation}\label{phi}
\dot \theta(s) = \cos(\varphi(s)), \quad \dot y = \sin(\varphi(s)).
\end{equation}
Note that these relations imply the identity $\dot\varphi(s)^2 = \ddot\theta(s)^2 + \ddot y(s)^2$. With this parametrization, the local curvature $\kappa^2(s) = \|\dot \vv(s)\|^2$ is given by
\[
\kappa^2(s) = \dot \theta(s)^4 + \ddot \theta(s)^2 +  \ddot y(s)^2 
= (\cos(\varphi(s)))^4+\dot \varphi(s)^2.
\]
In order to keep the notation as simple as possible, we shall from now on drop the explicit $s$-dependence of all functions.

\subsubsection*{The Euler-Lagrange equation}

The principle determining the MT's shape is the minimization of the total curvature over paths of fixed total length $L$, the length of the MT tip. The initial orientation of the MT, $\varphi(0) = \varphi_0$  (equivalently described by $\dot \theta(0)$ and $\dot y(0)$), is determined by the orientation locked in by the anchor at the base of the MT tip. Let 
\[
F_L(\varphi_0) = \{\varphi\in C^1([0,L];\mathbb{R}):\varphi(0) = \varphi_0\},
\]
We are looking for the minimizers of the functional
\begin{equation}
{\cal K}[\varphi] = \int_0^L ((\cos\varphi)^4+\dot \varphi^2)\; ds\qquad\varphi\in F_L(\varphi_0).
\end{equation}
(as given in the main text as Eq.~\eqref{eq:energyfunctional}.
To find the critical points of ${\cal K}$, we derive the associated \emph{Euler-Lagrange equation}. We consider variations \(\varphi(s) + r\xi(s)\) where $\xi\in F_L(0)$. The first variation in the direction of \(\xi\) is
\begin{align*}
\frac{d}{dr} {\cal K}[\varphi & +r\xi] \left.\vphantom{\frac{d}{dr}}\right|_{r=0} 
=  \int_0^L (-4(\cos\varphi)^3\sin\varphi \ \xi +2\dot \varphi\dot\xi)\;  ds\\ 
 & =  2\dot\varphi(L)\xi(L)  - 2\int_0^L \big( 2(\cos\varphi)^3\sin\varphi + \ddot \varphi\big) \xi\;\;  ds,
\end{align*}
where we have used $\xi(0)=0$ to eliminate one boundary term. If $\varphi$ is to be a critical point of $\cal K$, then this vanishes for all $\xi$, and in particular for all $\xi\in C^\infty_c((0,L))$ for which the first term vanishes. By the fundamental lemma of the calculus of variations, this implies that $\varphi$ is a solution of 
\begin{equation}
\frac{1}{2}\ddot \varphi + (\cos\varphi)^3\sin\varphi =0.
\label{eq:EL2}
\end{equation}
Once we know this, the first variation in the direction of $\xi$ reduces to $2\dot\varphi(L)\xi(L)$. This must vanish for all $\xi$ and hence $\dot\varphi(L)=0$.

In conclusion, $\varphi$ is a critical point for $\cal K$ if it solves \eqref{eq:EL2} with boundary conditions $\varphi(0)=\varphi_0$ and $\dot\varphi(L)=0$.

\section{Bifurcation analysis and buckling instability}
\label{sec:bucklingDetails}

For \(\varphi_0=0\), the constant \(\varphi(s)=0, p(s)=0\) is a solution in $F_L(0)$ for any value of \(L\). We seek to build a bifurcation diagram around this solution. First, we look for solutions having \(\varphi_0=0\) but such that \(p(0)>0\). If $\varphi_f = \varphi(L)$ is the value of the solution at the end point, the energy is given by
\begin{equation*}
E= H(\varphi(L),p(L))= - (\cos\varphi_f)^4
\end{equation*}
where we used that \(p(L)=0\) if a solution exists. In turn, this yields
\begin{equation}
p^2 =4 ((\cos\varphi)^4 - (\cos\varphi_f)^4),
\label{eq:p2}
\end{equation}
for all $s$. We now constrain ourselves to consider solutions that make a quarter rotation in the phase plane (starting on the line $\varphi=0$ and wrapping to the line $p=0$). For these, $L = \min\{s>0:p(s) = 0\}$. Furthermore, $p(s)>0$ for all $s<L$ so that~\eqref{eq:p2} and Hamilton's equation $2\dot\varphi = p$ yield
\begin{equation}
L = \int_{0}^{\varphi_f} \dfrac{1}{\sqrt{(\cos\varphi)^4-(\cos\varphi_f)^4}}d\varphi
\label{eq:Lofphif}
\end{equation}
This defines a function $L(\varphi_f)$ which gives the length of a MT tip required to ensure that the solution to equation \eqref{eq:cylsys} with boundary condition $\varphi(0)=0$ terminates at $\varphi(L)=\varphi_f$ when $p(L)=0$.
We prove the following three facts about $L$ as a function of $\varphi_f$ in Appendix \ref{sec:LphiProps}:
\begin{enumerate}
\item \(L\) is finite for \(0\le\varphi_f<\frac{\pi}{2}\).
\item \(L\) is monotone increasing on \([0,\pi/2)\), with \[\lim_{\varphi_f \to \frac{\pi}{2}} L(\varphi_f)= \infty.\]
\item For \(\varphi_f\) small, 
\begin{equation}
L(\varphi_f) \approx \frac{\pi}{2\sqrt2} + \frac{5\pi}{16\sqrt2} \varphi_f^2. \label{eq:LofPhifApprox}
\end{equation}
\end{enumerate}
In particular, 
\begin{equation*}
\min L = \frac{\pi}{2\sqrt{2}} =:L_0
\end{equation*}
proving the following bifurcation picture. If $L\le L_0$, the only solution with initial condition $\varphi_0=0$ is the steady state $\varphi= 0$. If however $L>L_0$, there exists another branch of solutions having $p(0)>0$ and as $L$ increases, the direction at the end of the MT tip continuously deflects further from the circumferential direction to the axial direction.

\begin{figure}
      \centering
      \includegraphics[width=8cm]{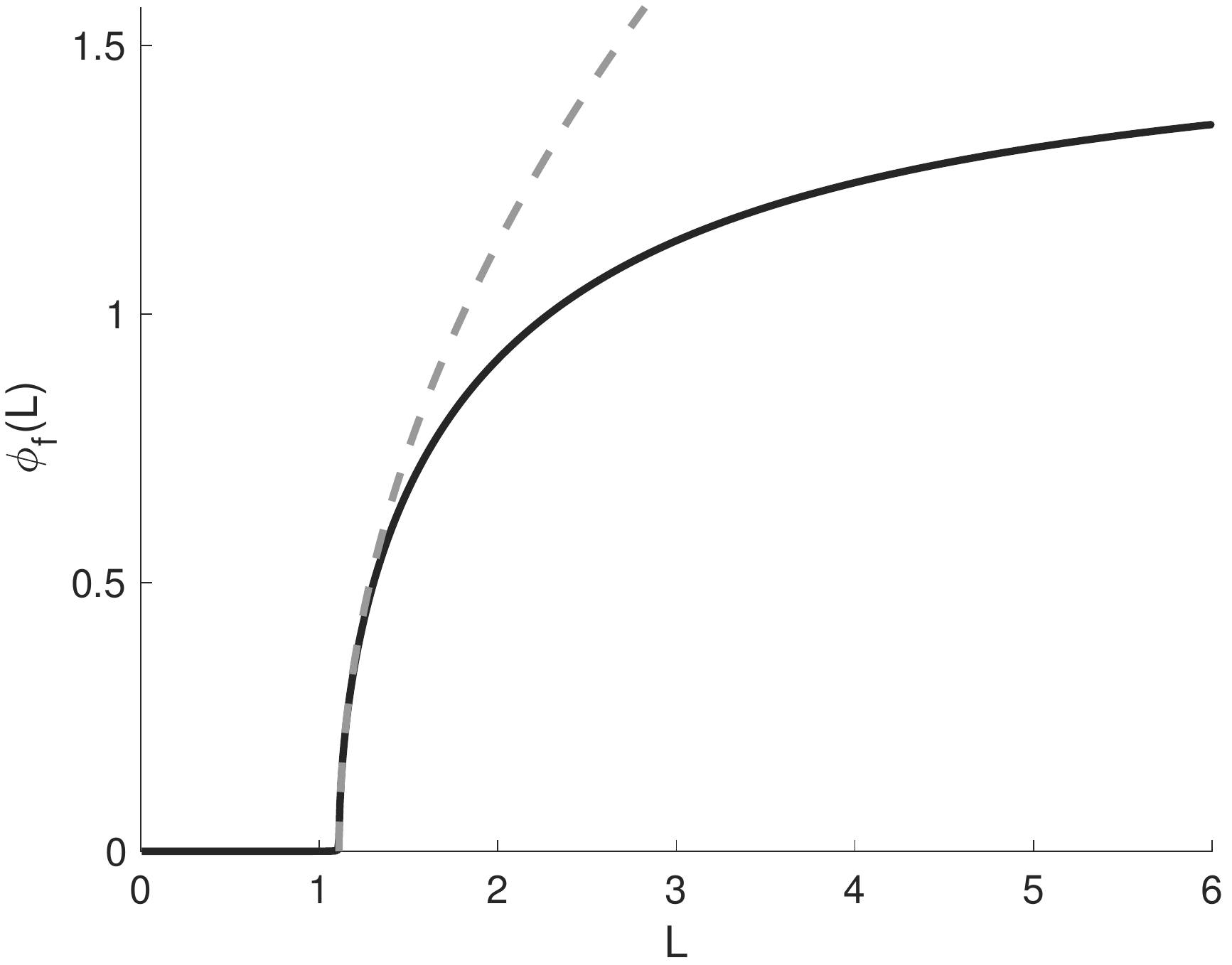} 

\caption{An approximate plot of \(\varphi_f(L)\) for $\varphi_0=0$ found by numerical integration of Eq \eqref{eq:Lofphif} (solid curve). The quadratic approximation is superimposed (grey, dashed). The shape of the pitchfork bifurcation shown here differs from the shape of the first pitchfork in Fig.~\ref{fig:pitchfork} because in Fig.~\ref{fig:pitchfork} we plot the solutions by $p(0)$ instead of $\varphi(L)$ for numerical convenience.}
\label{fig:phifofL}
\end{figure}

Inverting \(L(\varphi_f)\) gives us a characterization of this second solution described by its final value \(\varphi_f(L)\). Fig.~\ref{fig:phifofL} shows a numerical plot of \(\varphi_f(L)\) and the quadratic approximation given above.
A third solution is characterized by the other inverse branch \(-\varphi_f(L)\) and this corresponds to the symmetric solution in the phase plane starting at \(p(0)<0\). Combining the zero solution with these two spatially varying solutions in a single diagram, we see that the system seems to undergo a pitchfork bifurcation at \(L=\pi/(2\sqrt2)\).

In fact, let us show that there is a sequence of solutions bifurcating from the zero solution in the direction $\sin(\sqrt{2}s)$ for 
\begin{equation}
L_n = \frac{(1+2n)\pi}{2\sqrt{2}},n\in\mathbb{N},
\end{equation}
For this, it is best to rescale the variable by setting $s = rL$ and define
\begin{equation*}
\bar\varphi(r) = \varphi(Lr).
\end{equation*}
The curvature functional becomes
\begin{equation}
\bar{\cal K}_L[\bar\varphi] = \int_0^1\Big(L(\cos\bar\varphi)^4 + \frac{1}{L}(\bar\varphi')^2\Big)\:dr.
\label{eq:rescaledEL}
\end{equation}
While the Lagrangian now depends explicitly on $L$, the function space is $L$-independent, being $F=F_1(0)$ for all $L$. Now, the Euler-Lagrange equation can be cast as $G(L,\bar\varphi)=0$, where $G$ acts on the Banach space $\mathbb{R}\times F$. By the inverse function theorem, $G(L,\cdot)$ is invertible in a neighbourhood of the point $\varphi = 0$ as long as $G'(L,0)$ is invertible. Hence, it suffices to check that the kernel of $G'(L,0)$ is trivial. This is nothing else than showing that the linearized Euler-Lagrange equation
\begin{equation*}
\bar\varphi'' + 2L^2\bar\varphi = 0, \quad \bar\varphi(0)=0, \quad \bar\varphi'(L)=0
\end{equation*}
has no solution in $F$. The kernel of the operator $\partial^2 + 2L^2$ is the intersection of $F$ with the two-dimensional space of solutions of the differential equation, namely $\mathrm{span}\{\sin(\sqrt 2 s),\cos(\sqrt 2 s)\}$. It is non-trivial if and only if $\sqrt 2 L = \pi/2 + n \pi$ for any $n\in\mathbb{Z}$, in which case it is one-dimensional since the cosine never lies in $F$. Standard bifurcation arguments, see e.g.\ \cite{ambrosetti2007nonlinear}, then yield (i) neighbourhoods of $(L_n,0)$ in which all solutions are given by two branches, namely the trivial one and a continuous branch bifurcating in the direction of $\sin(\sqrt 2 s)$ and intersecting the trivial one only at $(L_n,0)$, and (ii) that there are no other bifurcation points along the zero solution. 

Let us finally discuss the stability of the zero solution. Note that $\bar{\cal K}_L[0] = L$. We claim that the zero solution is stable for $L<L_0$, by which we mean that 
\begin{equation*}
\bar{\cal K}_L[\bar\varphi] > \bar{\cal K}_L[0] = L
\end{equation*}
for all $L<L_0$ and all $\bar\varphi\neq 0$ in a neighbourhood of $0$ in $F$. Let us first observe that Eq.~\eqref{eq:rescaledEL} implies that for any fixed function $\bar\varphi\neq 0$, we have that $\bar{\cal K}_L[\bar\varphi]>L$ for $L$ small enough. Conversely, since $\int_0^1(\cos\bar\varphi)^4dr<1$ for any $\bar\varphi\neq 0$ the reverse inequality $\bar{\cal K}_L[\bar\varphi]<L$ holds for $L$ large enough. In fact, since
\begin{equation*}
\frac{d}{dL}(\bar{\cal K}_L[\bar\varphi] - L) 
= \int_0^1\big(((\cos\bar\varphi)^4 - 1)-\frac{1}{L^2}(\bar\varphi')^2\big)dr<0,
\end{equation*}
the function $L\mapsto \bar{\cal K}_L[\bar\varphi] - L$ is strictly decreasing. Hence, for each $\bar\varphi$ there is a unique $L(\bar\varphi)$ for which $\bar{\cal K}_{L(\bar\varphi)}[\bar\varphi] - L(\bar\varphi) = 0$. By the implicit function theorem, $L(\bar\varphi)$ is a continuous function. 

But then, the claim follows from the already established fact that there is no bifurcation point for $L<L_0$. Indeed, suppose that there is a path $[-1,0)\ni\sigma\mapsto\bar\varphi_\sigma$ such that $\bar\varphi_\sigma\to 0$ as $\sigma\to 0^-$ and such that $L(\bar\varphi_\sigma)\to L_*<L_0$ as $\sigma\to 0^-$. Then by construction $\bar{\cal K}_{L(\bar\varphi_\sigma)}(\bar\varphi_\sigma) = L(\bar\varphi_\sigma)\to L_* = \bar{\cal K}_{L_*}(0)$ and hence $\bar{\cal K}_{L_*}$ has a flat direction: there is a bifurcation at $(L_*,0)$, which is a contradiction.

\section{Properties of \texorpdfstring{\boldmath$L(\varphi_f)$}{L(phi)}}
\label{sec:LphiProps}

In order to simplify notation, we set $\varphi_f=x$. Setting $\varphi = xt$, we obtain
\[
L(x) = \int_0^1 \dfrac{x}{\sqrt{(\cos(tx))^4-(\cos x)^4}} dt.
\]

First of all, if $0<x<\frac{\pi}{2}$, the denominator is ${\cal O}((t-1)^{1/2})$ and hence integrable at $t=1^-$, so that $L$ is finite. If $x=\frac{\pi}{2}$ however, the integrand is of order $(t-1)^{-2}$ and hence not integrable. In order to show monotonicity, we compute 
\begin{equation*}
L'(x) = \int_0^1 \frac{G(xt)-G(x) }{((\cos(xt))^4-(\cos x)^4)^{3/2}} dt
\end{equation*}
where $G(x)=(\cos x)^4+2x(\cos x)^3\sin x$.
To show that $L'$ is positive it suffices to show that $G(xt)$ is decreasing for $t\in [0,1]$ for any $x\in[0,\pi/2]$.
The $t$ derivative of $G(xt)$ is $2x(\cos(xt))^2 (4tx(\cos(xt))^2-\sin(xt)\cos(xt)-3xt)$. Thus it suffices to show that
$4z(\cos z)^2-\sin z \cos z -3z \le 0$ for $z\in[0,\pi/2]$. The $z$ derivative of this expression is $-2\sin z (4z\cos z+\sin z)$ which is clearly negative in this range.

Finally, the small $x$ behaviour of Eq.~\eqref{eq:LofPhifApprox} is obtained from a Taylor expansion in x, with $t$ a parameter. Start with the Taylor expansion $(\cos x)^4=\sum_{n=0}^\infty a_n x^n$, where $a_0=1$, $a_2=-2$, $a_4=5/3$ and $a_n=0$ for $n$ odd. Using this we may write
\begin{align*}
\dfrac{(\cos(tx))^4-(\cos x)^4}{x^2} & = \\
 2(1-t^2) & \left(
1 - (5/6)(t^2+1)x^2 +  \sum_{n=6}^\infty a_n\dfrac{t^n-1}{1-t^2} x^{n-2}
\right)
\end{align*}
The coefficients $a_n$ arising from the expansion of an entire function are exponentially decreasing in $n$ so it is not hard to see that $\sum_{n=6}^\infty a_n\dfrac{t^n-1}{1-t^2} x^{n-2} = O(x^4)$, uniformly for $t\in [0,1]$. Raising both sides to the power $-1/2$ and using the expansion $(1+x)^{-1/2} =1-(1/2)x + O(x^2)$ leads to 
\begin{align*}
\left(\dfrac{(\cos(tx))^4-(\cos x)^4}{x^2}\right)^{-1/2} &\\
 = (2(1-t^2))^{-1/2} & \left(
1 + (5/12)(t^2+1)x^2 + O(x^4)
\right).
\end{align*}
The result now follows from integrating with respect to $t$.

\section{Proof of a bound on \texorpdfstring{$\Delta_L$}{Delta L}}
\label{sec:DeltaBoundProof}

Here we prove inequality~\eqref{eq:DeltaBound} which stated that $\vert \Delta_L(\varphi_0,n) \vert\leq \frac{L^2}{n}$. To show this, we note that the Euler-Lagrange equation~\eqref{eq:cylsys} immediately implies that $\vert\dot p\vert\leq 2$ and hence
\begin{equation*}
\vert p(s)\vert\leq \int_s^{\ell}\vert\dot p(r)\vert dr \leq 2\ell,
\end{equation*}
where we used that $p(\ell)=0$. By the dynamical equation again, we thus obtain the uniform bound
\begin{equation*}
\vert \varphi_j(\ell) - \varphi_j(0) \vert 
\leq \frac{1}{2}\int_0^{\ell}\vert p(r) \vert dr
\leq\ell^2.
\end{equation*}
Hence, the total angular change is given by
\begin{equation*}
\left\vert\Delta_L(\varphi_0,n)\right\vert
= \Big\vert\sum_{j=1}^{n}(\varphi_j(L/n) - \varphi_j(0))\Big\vert
\leq n\ell^2 = \frac{L^2}{n},
\end{equation*}
where we recalled that $\ell = L/n$. This establishes the bound we had set out to show.

\section*{Acknowledgments}
The authors would like to thank Geoff Wasteneys for useful feedback on the manuscript. This work was supported by the National Science and Engineering Research Council of Canada.

\bibliographystyle{siamplain}
\bibliography{PlantMTs2}
\end{document}